\documentclass[prb,showpacs,footinbib,twocolumn,final,superscriptaddress]{revtex4-1}

\usepackage{amsfonts,amsmath,amssymb,ifpdf,epsfig,graphicx}
\usepackage[usenames,dvipsnames]{color}
\usepackage{hyperref}
\usepackage{bm}

\newcommand{\vq}{{\bm{q}}}

\newcommand{\vx}{{\bm{x}}}
\newcommand{\vy}{{\bm{y}}}
\newcommand{\vj}{{\bm{j}}}
\newcommand{\vE}{{\bm{E}}}

\newcommand{\wo}{\omega}

\newcommand{\dx}[1]{ \! {\rm d}#1 \, }
\newcommand{\dint}{\int}

\newcommand{\Eq}[1]{Eq.~\eqref{#1}}
\newcommand{\EqPlain}[1]{\eqref{#1}}
\newcommand{\EqsFromTo}[2]{Eqs.~(\ref{#1})--(\ref{#2})}
\newcommand{\EqsOne}[1]{Eqs.~(\ref{#1})}

\newcommand{\Fig}[1]{Fig.~\ref{#1}}
\newcommand{\Ref}[1]{Ref.~[\onlinecite{#1}]}
\newcommand{\Refs}[1]{Refs.~[\onlinecite{#1}]}

\newcommand{\Ima}[1]{{\rm Im} \! \left[#1\right] }
\newcommand{\Rea}[1]{{\rm Re} \! \left[#1\right] }

\newcommand{\RPA}{{\rm RPA}}

\newcommand{\typ}{{\rm typ}}

\newcommand{\inter}{{\rm int}}

\newcommand{\on}{\hat{n}}
\newcommand{\oV}{\hat{V}}
\newcommand{\op}{\hat{\psi}}
\newcommand{\opd}{\hat{\psi}^{\dagger}}

\newcommand{\CFV}{\langle |V|^2 \rangle}
\newcommand{\CFj}{\langle |\vj|^2 \rangle}
\newcommand{\CFn}{\langle |n|^2 \rangle}

\newcommand{\ichi}{\Upsilon}
\newcommand{\emfp}{\ell}
\newcommand{\clasres}{\mathcal{R}}
\newcommand{\vertexm}{\mathcal{M}}

\newcommand{\splus}{\!+\!}

\newcommand{\mean}[1]{\left\langle #1 \right\rangle}
\newcommand{\smean}[1]{\langle #1 \rangle}


\begin{document}

\title{Thermal noise and dephasing due to electron interactions in
       non-trivial geometries}

\author{M.~Treiber}
\affiliation{Ludwig Maximilians University, Arnold Sommerfeld Center
             and Center for Nano-Science, Munich, D-80333, Germany}

\author{C.~Texier}
\affiliation{Univ. Paris Sud ; CNRS ; LPTMS, UMR 8626 \& LPS, UMR 8502,
             Orsay F-91405, France}

\author{O.~M.~Yevtushenko}
\affiliation{Ludwig Maximilians University, Arnold Sommerfeld Center
             and Center for Nano-Science, Munich, D-80333, Germany}

\author{J.~von~Delft}
\affiliation{Ludwig Maximilians University, Arnold Sommerfeld Center
             and Center for Nano-Science, Munich, D-80333, Germany}

\author{I.~V.~Lerner}
\affiliation{School of Physics and Astronomy, University of Birmingham,
             Birmingham, B15 2TT, UK}

\date{\today}

\begin{abstract}
We study Johnson-Nyquist noise in macroscopically inhomogeneous
disordered metals and give a microscopic derivation of the correlation
function of the scalar electric potentials in real space. Starting from
the interacting Hamiltonian for electrons in a metal and the random
phase approximation, we find a relation between the correlation function
of the electric potentials and the density fluctuations which is valid
for arbitrary geometry and dimensionality.
We show that the potential fluctuations are proportional to the solution
of the diffusion equation, taken at zero frequency. As an example, we
consider networks of quasi-1D disordered wires and give an explicit
expression for the correlation function in a ring attached via arms to
absorbing leads. We use this result in order to develop a theory of
dephasing by electronic noise in multiply-connected systems.
\end{abstract}

\maketitle

\section{Introduction}
\label{SectIntroduction}

Electronic noise generated by the thermal excitation of charge carriers
has been observed and explained by Johnson and Nyquist more than 80
years ago \cite{JohnsonNyquist_ThermalAgitation_1928} and discussed in
great detail in the literature since then. More recently, it has been
found that this so-called  Johnson-Nyquist noise is the main source of
dephasing in mesoscopic systems at low temperatures of a few Kelvins,
where phonons are frozen out. Dephasing puts an IR cut-off for
interference phenomena, such as quantum corrections to the classical
conductivity.\cite{AAK_Localization_1982}

\begin{figure}[t]
  \centering
	\includegraphics{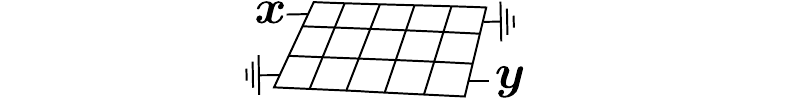}
	\caption{A network of wires: We are interested in the noise
	         correlations between arbitrary points $\vx$ and $\vy$ of
	         multiply-connected networks attached to leads (denoted by
	         the usual ground symbol) at arbitrary points.}
	\label{FigGraph}
\end{figure}

The current interest in this topic arises from studies of dephasing in
mesoscopic systems which consist of connected quasi-1D disordered wires,
\Fig{FigGraph}, including connected rings and grids.
\cite{Ferrier_Networks_2004,Schopfer_Networks_2007} It has been found
(both experimentally \cite{Ferrier_Networks_2008} and theoretically
\cite{Ludwig_Ring_2004,Texier_Ring_2004,Texier_Multiterminal_2004,
Treiber_Dephasing_2009,Treiber_Book_2010}) that dephasing depends not only on the
dimensionality, but also on the geometry of the system. The noise
correlation function is well-understood for macroscopically homogeneous
systems such as infinite wires or isolated rings, but has so-far not
been studied in multiply-connected networks with leads attached at
arbitrary points. The goal of this paper is to give a transparent and
systematic description of the thermal noise properties for such systems.
In particular, we will derive an expression for the fluctuations of the
scalar electric potentials for arbitrary geometries, \Eq{FinalCorrFunc},
and a general expression for the corresponding dephasing rate,
\Eq{eq:FunctionalDecoh}. Throughout, we assume that a description of the
noise in terms of scalar potentials is sufficient, i.e. we neglect the
fluctuations of the transverse component of the electromagnetic field
(for a detailed discussion of the latter, see
\Ref{AAK_Localization_1982}).

Let us start by reviewing simplified arguments to derive the noise
correlation function: Johnson and Nyquist concluded that thermal noise
in electrical conductors is approximately white, meaning that the power
spectral density is nearly constant throughout the whole frequency
spectrum. If in addition the fluctuations are uncorrelated for different
points in space, a correlation function for the random thermal currents
in the classical limit is independent of frequency $\wo$ and momentum
$\vq$. The power spectrum of the current density reads
\begin{equation}\label{Noisej}
	\CFj(\vq,\wo) = 2 T \sigma_0 \, .
\end{equation}
Here, $\sigma_0 = e^2 \nu D $ is the Drude conductivity of the
disordered system, $D$ and $\nu$ are the diffusion constant and the
density of states, respectively.
Naively applying Ohm's law, $\vj(\vq) = \sigma_0 \vE(\vq)$, to
\Eq{Noisej} and using the relation between the electric field and the
scalar potential, $e \vE(\vq) = - i \vq V(\vq)$, we find
\begin{equation}\label{VVAAK}
	\CFV(\vq,\wo)
	= \frac{2 T e^2}{\sigma_0} \frac{1}{\vq^2} \, .
\end{equation}
The correlation function, \Eq{VVAAK}, corresponds to the coupling of a
given electron to the bath of the surrounding electrons.
\cite{AAK_Localization_1982} Thus $\CFV$ describes the process of
successive emission and re-absorption of a photon, which is described
effectively by the scalar potential $V$. The factor $1/\vq^2$ coincides
with the solution of a diffusion equation in an infinite system, which
reflects the fact that the currents, \Eq{Noisej}, are uncorrelated in
space.

These simple arguments are based on the homogeneity of the system and
have assumed a local relation between potential and current, whereas
transport properties in disordered metals are substantially non-local.
\cite{Zyuzin_Fluctuations_1987,Kane_Correlations_1988,
Lerner_CurrentDistribution_1989,Texier_Multiterminal_2004}
In this paper, we derive an analogy of \Eq{VVAAK} for disordered systems
with arbitrary geometry and dimensionality; this will in particular
apply to networks of disordered wires. A detailed calculation, which
takes into account all properties of the mesoscopic samples, has to be
done in the real-space representation. Starting points are the usual
linear response formalism and the fluctuation-dissipation theorem (FDT).
\cite{Landau_Book_1980} Although most ingredients of the following
discussion will be familiar to experts, we hope that the manner in which
they have been assembled here will be found not only to be pedagogically
useful, but also helpful for further theoretical studies.

The paper is organized as follows: In Section \ref{Heuristic}, we
propose a heuristic description of the potential fluctuations. In
Section \ref{SectNoiseCorr}, we review a microscopic approach to the
noise correlation function, based on a relation of the fluctuations of
the scalar potentials to the fluctuations of the density, using the
random phase approximation (RPA). In Section \ref{SectDensityResponse}
we evaluate the density response function $\chi$ for disordered
systems by using a real-space representation for arbitrary geometries.
We apply this result to the noise correlation function in Section
\ref{SectFinalCorrFunc}. In Section \ref{SectNetworks}, we show how the
noise correlation function can be calculated for networks of disordered
wires. Finally, in Section \ref{Dephasing} we discuss the relation to
the fundamental problem of dephasing by electronic interactions.

\section{Heuristic description of potential fluctuations}
\label{Heuristic}

A description of fluctuations in metals within the linear response
formalism naturally starts with an analysis of the density fluctuations
in the model of non-interacting electrons described by the standard
free-electron Hamiltonian $\hat{H}^{(0)}$. This system is perturbed by
an external scalar potential $V(\vx,t)$ coupled to the density operator
$\on(\vx)$:
\begin{equation}\label{defHInt1}
	\hat{H}^{(1)} =  \int \dx{\vx} V(\vx,t) \, \on(\vx) \, .
\end{equation}
The response of the (induced) charge density,
\begin{align}
	n_\mathrm{ind}(\vx,\wo)
        &\equiv
         \int_{-\infty}^{\infty}\dx{t} e^{i\wo t}
          \left[\mean{ \on(\vx,t) }_\mathrm{pert} - \mean{ \on(\vx) } \right]
        \nonumber\\
	&= - \int \dx{\vy} \chi(\vx,\vy,\wo) \, V(\vy,\wo) \, ,
  \label{LinResponseV}
\end{align}
is governed by the (retarded) density response function:
\begin{align}
  \label{DefChi}
	\chi(\vx,\vy,\wo)
	& = i \int_0^{\infty} \dx{t} e^{i(\wo+i0)t}
		\mean{ [\on(\vx,t),\on(\vy,0)] } \, .
\end{align}
Here $\mean{\cdots}_\mathrm{pert}$ and $\mean{\cdots}$ denote
quantum/statistical averaging with respect to the perturbed and
unperturbed Hamiltonian, respectively. The FDT relates the equilibrium
density fluctuations to the imaginary (dissipative) part of the response
function,
\begin{align}
  \label{nnChi}
  \CFn(\vx,\vy,\wo)
  &\equiv\int_{-\infty}^{\infty}\dx{t} e^{i\wo t}
		\mean{ \on(\vx,t) \, \on(\vy,0) } \\
  \label{EqFDTnn}
  &= F(\wo)\, \Ima{\chi(\vx,\vy,\wo)} \, ,
\end{align}
where
\begin{equation}\label{EqFDTF}
	F(\wo) = \frac{2}{1-e^{-\wo/T}} \, .
\end{equation}
In writing \EqsFromTo{nnChi}{EqFDTF} we have exploited detailed balance
and time-reversal symmetry. The latter implies
$\chi(\vx,\vy,\wo)=\chi(\vy,\vx,\wo)$.\cite{Landau_Book_1980}

The question which we are going to address in this paper is how to
characterize the fluctuations of the electric potential $V$. For
this purpose we consider the ``dual'' case, where some external density
$n_\mathrm{ext}(\vx,t)$ is the perturbation that couples to the
``potential operator'' $\oV$:\cite{Footnote_VOperator}
\begin{equation}\label{defHInt2}
	\hat{H}^{(2)} =  \int \dx{\vx} \oV(\vx) \, n_\mathrm{ext}(\vx,t) \, .
\end{equation}
The linear response of $\oV$ to the perturbation can be written as
\begin{equation}\label{LinResponsen}
	\smean{ \oV(\vx,\wo) }_\mathrm{pert}
	= \int \dx{\vy} \ichi(\vx,\vy,\wo) \, n_\mathrm{ext}(\vy,\wo) \, ,
\end{equation}
defining the response function $\ichi$. In analogy to \Eq{EqFDTnn}, the
response function also characterizes the equilibrium fluctuations of the
potential:\cite{Footnote_Sign}
\begin{equation}\label{VVIChi}
	\CFV(\vx,\vy,\wo)
	= F(\wo) \, \Ima{-\ichi(\vx,\vy,\wo)} \, .
\end{equation}

Calculating the response function $\ichi(\vx,\vy,\wo)$ is a complicated
task because it requires precise knowledge of the potential operator
$\oV(\vx)$. Instead, we can identify the potential $V(\vx,\wo)$ in
\Eq{LinResponseV} with the response
$\smean{ \oV(\vx,\wo) }_\mathrm{pert}$ in \Eq{LinResponsen} to relate
$\ichi$ to $\chi$: In the limit of strong screening in good conductors
(called the unitary limit~\cite{Treiber_Book_2010}), electroneutrality
is satisfied locally. Therefore, the induced charge exactly compensates
the external charge: $n_\mathrm{ind}(\vx,\wo)=-n_\mathrm{ext}(\vx,\wo)$.
Now inserting \Eq{LinResponseV} into \Eq{LinResponsen} (or vice versa),
we obtain
\begin{equation}\label{IChiChi}
  \int \dx{\vx'} \ichi(\vx,\vx',\wo) \chi(\vx',\vy,\wo)
  = \delta(\vx-\vy) \, .
\end{equation}
If $\chi$ is known, \EqsOne{VVIChi} and \EqPlain{IChiChi} allow one to
calculate the correlation function of the scalar potential.

Let us recall the well-known case of macroscopically homogeneous
diffusive systems. The expression for the disordered averaged response
function $\overline{\chi}$ reads\cite{Vollhardt_Localization_1980,
Akkermans_Book_2007}
\begin{equation}
  \label{eq:ChiDiffusiveFourier}
  \overline{\chi}(\vq,\wo)
  = \nu \frac{D\vq^2}{D\vq^2 - i\wo}
  = 1/\ichi(\vq,\wo) \, ,
\end{equation}
where we used \Eq{IChiChi}. Inserting \Eq{eq:ChiDiffusiveFourier} into
\Eq{VVIChi}, we find
\begin{equation}
  \CFV(\vq,\wo)
  = \frac1\nu \frac{\wo F(\wo)}{D\vq^2} \, ,
\end{equation}
which reduces to \Eq{VVAAK} in the limit $\wo \ll T$.

\section{Noise correlation function for arbitrary geometries: microscopic approach}
\label{SectNoiseCorr}

In \EqsOne{defHInt1} and \EqPlain{defHInt2}, we introduced the operators
$\on$ and $\oV$ assuming that either $V(\vx,t)$ or
$n_\mathrm{ext}(\vx,t)$ are external perturbations. In fact, the
fluctuations originate inside of the system and the starting point of a
microscopic description is the part of the Hamiltonian, which describes
electron interactions,
\begin{equation}\label{Hint}
  \hat{H}_\inter = \dint \dx{\vx} \dx{\vy}
            U_0(\vx,\vy) \opd(\vx) \opd(\vy) \op(\vy) \op(\vx) \, ,
\end{equation}
where $U_0(\vx,\vy)$ is the bare Coulomb interaction. In the mean-field
approximation, \Eq{Hint} gives rise to a correction, called Hartree
contribution, to the electron energy:
\begin{equation}\label{HartreeContrib}
	\Delta E^{(\mathrm{Hartree})}
	   \approx \dint \dx{\vx} \dx{\vy} U_0(\vx,\vy)
	   \langle \on(\vx) \rangle \langle \on(\vy) \rangle
	   \, ,
\end{equation}
where $\on(\vx)=\opd(\vx)\op(\vx)$.

\begin{figure}[t]
  \centering
	\includegraphics{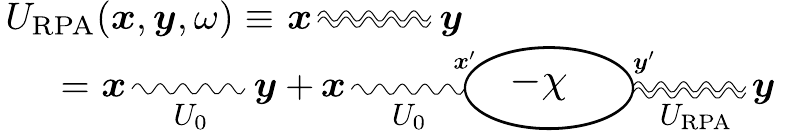}
	\caption{The Coulomb interaction in the RPA according to
	         \Eq{EqInitialIntEq}.
	        }
	\label{FigURPA}
\end{figure}
The Coulomb interactions are dynamically screened, which can be
accounted for in the framework of the RPA, provided that the electron
density is high,
\begin{align}\nonumber
	& U_{\RPA}(\vx,\vy,\wo)
	= U_0(\vx,\vy) - \dint \dx{\vx'}\dx{\vy'} U_0(\vx,\vx') \chi(\vx',\vy',\wo)
	\\ \label{EqInitialIntEq}
	& \qquad \qquad \qquad \qquad \times  U_{\RPA}(\vy',\vy,\wo) \, ,
\end{align}
see \Fig{FigURPA}. Note that $-\chi$ [see the definition in \Eq{DefChi}]
is equal to the bubble diagrams of \Fig{FigURPA}, see e.g.
\Ref{Akkermans_Book_2007}. In Appendix \ref{AppendixScreening}, we
recall how to obtain \Eq{EqInitialIntEq} within a self consistent
treatment of the screening problem.

Using the RPA in \Eq{HartreeContrib} and comparing the result
with equation \Eq{defHInt2}, we observe that the potential fluctuations
are due to electronic interactions and that the operator of the scalar
potential is given by
\begin{equation}\label{defVfunc}
  \oV(\vx,\wo) = \int \dx{\vy} U_{\RPA}(\vx,\vy,\wo)\, \on(\vy) \, .
\end{equation}

\begin{figure}[t]
  \centering
	\includegraphics{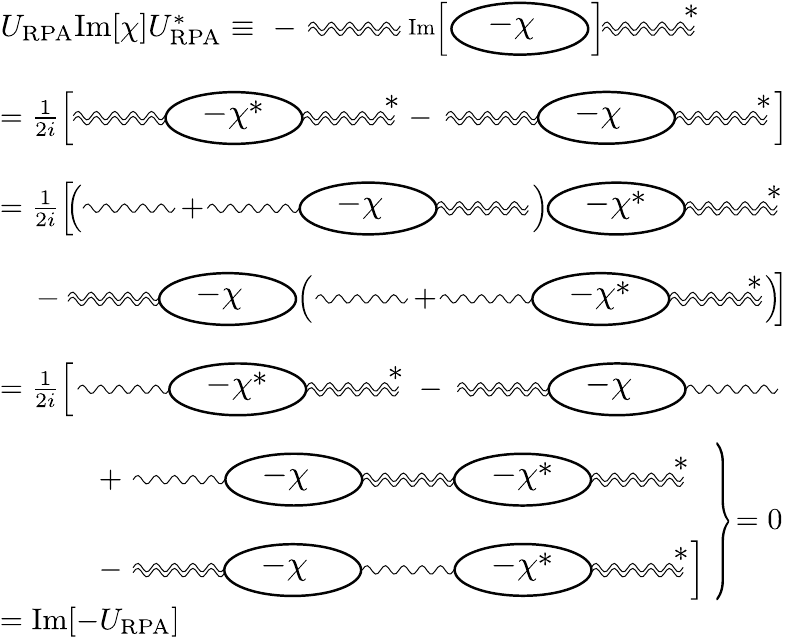}
	\caption{Diagrammatic proof of \Eq{EqFDTVV} by using \Eq{EqDefVV}
	         and \Eq{EqInitialIntEq} (i.e. \Fig{FigURPA}).
          }
	\label{ProofURPA}
\end{figure}

\Eq{defVfunc} allows us to relate the correlation function of the
potentials to the correlation function of the density fluctuations:
\begin{align}\nonumber
	& \CFV(\vx,\vy,\wo)
	= \dint \dx{\vx'} \dx{\vy'} U_{\RPA}(\vx,\vx',\wo)
	\\ \label{EqDefVV}
	& \qquad \qquad \qquad \quad \ \times \CFn (\vx',\vy',\wo)
	U_{\RPA}^*(\vy',\vy,\wo) \, .
\end{align}
By inserting \EqsOne{EqInitialIntEq} and \EqPlain{EqFDTnn} into
\Eq{EqDefVV}, re-ordering the terms in the RPA series and using the fact
that $U_0$ is real, we find (see \Fig{ProofURPA})
\begin{equation}\label{EqFDTVV}
	\CFV(\vx,\vy,\wo)
	= F(\wo) \, \Ima{-U_{\RPA}(\vx,\vy,\wo)} \, .
\end{equation}
We emphasize that the derivation of \Eq{EqFDTVV} has not used any other
assumption than the RPA. Thus, \EqsOne{EqInitialIntEq} and \EqPlain{EqFDTVV} are a
microscopic (and more rigorous) counterpart of the phenomenological
\EqsOne{VVIChi} and \EqPlain{IChiChi}.

\section{Density response in disordered systems: calculations in coordinate representation}
\label{SectDensityResponse}

\begin{figure}[t]
	\includegraphics{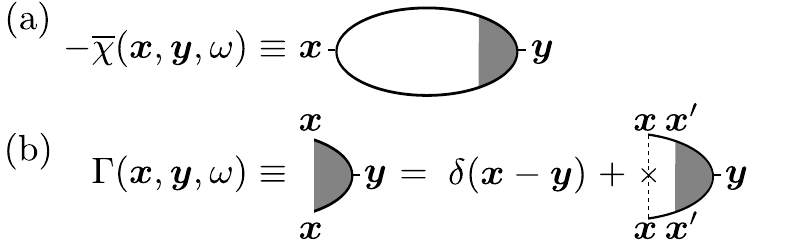}
	\caption{(a) Equation for the disorder averaged density response
	         function; solid lines denote the disorder averaged
	         retarded and/or advanced Green functions, cf. \Eq{ChiGG}.
	         (b) Equation for the impurity vertex; the dashed line
	         represents impurity scattering, cf. \Eq{DefGamma}.}
  \label{FigChi0v}
\end{figure}

In disordered metals the motion of the electrons is  diffusive, provided
that $k_F^{-1}\ll \emfp \ll L$ where $k_F$ is the Fermi wave-vector,
$\emfp$ the mean free path and $L$ the system size.
It can be accounted for by substituting the disorder-averaged density
response function, $\overline{\chi}$, into the phenomenological
\EqsOne{VVIChi} and \EqPlain{IChiChi} or the microscopic
\EqsOne{EqInitialIntEq} and \EqPlain{EqFDTVV}. The function $\overline{\chi}$ has been calculated
for macroscopically homogeneous systems by Vollhardt and W\"olfle.
\cite{Vollhardt_Localization_1980} In the following, we will show how to
generalize their calculation to inhomogeneous systems. A useful starting
point is a coordinate representation of the density response function,
\Eq{DefChi}, in terms of the advanced and retarded Green's functions,
$G^{R/A}(\vx,\vy,\wo)$,
\begin{align}\label{ChiGG}
  \overline{\chi}(\vx,\vy,\wo)
  = & \frac{1}{2 \pi i} \int \dx{\epsilon} \Big( \left[ f(\epsilon+\wo) - f(\epsilon) \right]
      \\ \nonumber
   & \times
      \overline{G^R(\vx,\vy,\epsilon+\wo)G^A(\vy,\vx,\epsilon)}
      \\ \nonumber
  & + f(\epsilon)
      \overline{G^R(\vx,\vy,\epsilon+\wo)G^R(\vy,\vx,\epsilon)}
      \\ \nonumber
 & - f(\epsilon+\wo)
      \overline{G^A(\vx,\vy,\epsilon+\wo)G^A(\vy,\vx,\epsilon)} \Big)
  \, ,
\end{align}
see Fig.~\ref{FigChi0v}(a). Here, $f(\wo)$ is the Fermi distribution
function, and $\overline{\cdots}$ denotes disorder averaging.
The combinations $\overline{G^R G^R}$ and $\overline{G^A G^A}$ give
short-range contributions, since the average of the products decouple,
e.g. $\overline{G^R G^R} \simeq \overline{G^R} \cdot \overline{G^R} +
{\cal O}(1/k_F\emfp)$, and the disorder averaged Green's functions
$\overline{G^R}$ and $\overline{G^A}$ decay on the scale $\emfp \ll L$.
We will consider contributions to the thermal noise which are governed
by distances larger than $\emfp$, cf. \Ref{AAK_Localization_1982}.
Therefore, details of the behavior on short scales are not important for
our purposes and we replace the short-range contributions,
$\overline{G^R G^R}$ and $\overline{G^A G^A}$, by a delta function.
The long range contributions $\overline{G^R G^A}$ can be calculated by
standard methods,\cite{Akkermans_Book_2007}
\begin{align}\nonumber
	& \overline{G^R(\vx,\vy,\epsilon+\wo)G^A(\vy,\vx,\epsilon)}
	= \int \dx{\vx'} \overline{G^R}(\vx,\vx',\epsilon+\wo)\\
	 & \qquad\qquad\qquad\qquad\qquad   \times \overline{G^A}(\vx',\vx,\epsilon)
	    \Gamma(\vx',\vy,\wo) \, ,
\end{align}
where $\Gamma(\vx,\vy,\wo)$ is the impurity vertex function,
\begin{align}\nonumber
   \Gamma(\vx,\vy,\wo)
  &= \delta(\vx-\vy) 
    +\frac1{2\pi \nu \tau}
    \!\! \int \dx{\vx'}
    \overline{G^R}(\vx,\vx',\epsilon+\wo)\\ \label{DefGamma}
  & \qquad \times  \overline{G^A}(\vx',\vx,\epsilon)
    \Gamma(\vx',\vy,\wo)
\end{align}
(the factor $1 / 2 \pi \nu \tau$, where $\tau = \emfp/v_F$ is the
transport time, originates from the impurity line), see
Fig.~\ref{FigChi0v}(b). The short-ranged product,
$\overline{G^R}\cdot\overline{G^A}$, can be expanded as
$\overline{G^R}\cdot\overline{G^A}\simeq2\pi \nu \tau\,
\delta(\vx-\vx')\big[1+i\wo\tau+\tau D\Delta_{\vx}\big]$,
which is obtained by transforming the product to momentum space and expanding
in the transferred momentum $\vq$ and frequency $\wo$, realizing that terms of order $\vq$
vanish due to symmetry.
As a result, \Eq{DefGamma} reduces to a diffusion equation:
\begin{equation}\label{GammaDifEq}
  \left(-i\wo -D\Delta_{\vx} \right) \Gamma(\vx,\vy,\wo)
  = \frac{1}{\tau} \delta(\vx-\vy) \, ,
\end{equation}
where $D =v_F\emfp/d$ is the diffusion constant for a $d$ dimensional
system. Thus, the vertex
function is proportional to the diffusion propagator,
$\Gamma(\vx,\vy,\wo) = P(\vx,\vy,\wo) / \tau$.

Collecting the short- and long-range contributions and taking the limit
$T \ll \epsilon_F$, we obtain from \Eq{ChiGG}
\begin{equation}\label{ChiBar}
  \overline{\chi}(\vx,\vy,\wo)  
  = \nu \left( \delta(\vx-\vy) + i \wo P(\vx,\vy,\wo) \right) \, .
\end{equation}
\Eq{ChiBar} is valid for arbitrary geometries since it is based only on
the diffusive approximation and does not require macroscopic
homogeneity.

\section{Noise correlation function in disordered systems}
\label{SectFinalCorrFunc}

Let us simplify \Eq{EqInitialIntEq} for a disordered conductor. Using
\Eq{ChiBar} and
\begin{equation}\label{EqPoisson}
	-\frac{1}{4 \pi e^2} \Delta_\vx U_0(\vx,\vx') = \delta(\vx-\vx') \, ,
\end{equation}
\Eq{EqInitialIntEq} can be written as
\begin{align}\nonumber
	& \left( 1-\frac{\Delta_\vx}{\kappa^2} \right)
	  U_{\RPA}(\vx,\vy,\wo) +i\wo \int \dx{\vx'} P(\vx,\vx',\wo)  \\ \label{EqFinalIntEq}
	& \quad \times U_{\RPA}(\vx',\vy,\wo)
	  = \frac{1}{\nu}\delta(\vx-\vy) \, ,
\end{align}
where we introduced the Thomas-Fermi screening wave-vector
$\kappa = \sqrt{4 \pi e^2 \nu}$, which corresponds to the inverse
screening length in three dimensional (3D) bulk systems. The kernel of \Eq{EqFinalIntEq} is
a solution to the diffusion equation \EqPlain{GammaDifEq} which can be
expanded in terms of eigenfunctions of the Laplace operator.
Consequently, the kernel is always separable and \Eq{EqFinalIntEq} has a
unique solution (see e.g. \Ref{Kanwal_IntegralEquations_1997} for
details on how the solution can be found). Using the semi-group property
of the diffusion propagators,
\begin{equation}\nonumber
	\int \dx{\vx'} P(\vx,\vx',\wo) P(\vx',\vy,0)
	 = \frac{i}{\wo} \left[ P(\vx,\vy,0) - P(\vx,\vy,\wo) \right] \, ,
\end{equation}
one can check that
\begin{equation}\label{ResultURPA}
	U_{\RPA}\!(\vx,\vy,\wo)
	\!=\! \frac{1}{\nu} \!\left(\! \frac{1}{-D\Delta_\vx -i\wo}
	\!+\! \frac{1}{D\kappa^2} \!\right)^{\!\!\!-1} \!\!\! P(\vx,\vy,0)
\end{equation}
satisfies \Eq{EqFinalIntEq}. In practice, the 3D Thomas-Fermi screening
length $\kappa^{-1}$ is a microscopic scale, thus the typical value of
the first term of the rhs. of \Eq{ResultURPA},
$(D\vq^2_{\typ} - i\wo_{\typ})^{-1}$, is larger than
$1/D\kappa^2=1/4\pi\sigma_0$ for good conductors (this is the so-called
unitary limit, for details see \Ref{Treiber_Book_2010}):
\begin{equation}\label{ValidityUnitaryLimit}
	\frac{1}{|D\vq^2_{\typ} - i\wo_{\typ}|}
	\gg \frac{1}{D\kappa^2} \, .
\end{equation}
In this limit, using the diffusion \Eq{GammaDifEq}, we obtain from \Eq{ResultURPA}:
\begin{equation}\label{URPADiff}
    U_{\RPA}(\vx,\vy,\wo)
    = \frac{1}{\nu} \left[ \delta(\vx-\vy) - i\wo P(\vx,\vy,0) \right] \, .
\end{equation}
We remind that $P(\vx,\vy,0)$ is always real. As a result, \EqsOne{EqFDTVV} and \EqPlain{URPADiff} yield
\begin{equation}\label{FinalCorrFunc}
	\CFV(\vx,\vy,\wo)
	= \frac{1}{\nu} \, \wo F(\wo) P(\vx,\vy,0) \, ,
\end{equation}
where $F(\wo)$ is given by \Eq{EqFDTF}.
The real-space demonstration of \EqsOne{URPADiff} and \EqPlain{FinalCorrFunc}
for macroscopically inhomogeneous systems, are among the main results
of the paper. It is worth emphasizing the frequency-space factorization
of the correlator, which plays an important role in the theory of
dephasing, cf. Section~\ref{Dephasing}.
The relation of \Eq{FinalCorrFunc} to the correlation function of the
currents, \Eq{Noisej}, is discussed in Appendix
\ref{appendix:CurrentCorrelation}, and allows to put the
presentation of the introduction on firm ground.

Note that \Eq{ValidityUnitaryLimit} allows one to neglect the term
$\Delta_\vx/\kappa^2$ in \Eq{EqFinalIntEq} and, thus, to reduce
\Eq{EqFinalIntEq} to the form of the phenomenological integral equation
\EqPlain{IChiChi}, with $U_{\RPA}$ taking the place of $\ichi$.
[The same replacement leads from \Eq{VVIChi} to \Eq{EqFDTVV}.]
In other words, the electric potential of the fluctuating charge
densities itself is negligible when screening is strong enough (i.e.
good conductors in the unitary limit), justifying a \emph{posteriori} our
assumptions in the phenomenological Section \ref{Heuristic}.

The fact that the correlation function of the potential is proportional
to the solution of the diffusion equation at zero frequency, cf.
\Eq{FinalCorrFunc}, may be understood as a non-local version of the
Johnson-Nyquist theorem, since $P(\vx,\vy,0)$ can be related to the
classical dc-resistance $\clasres(\vx,\vy)$ between the points $\vx$ and
$\vy$ (see \Ref{Texier_NetworksCylinders_2009}):
\begin{equation}
  \label{Resistance}
  \clasres(\vx,\vy)=\frac{2D}{\sigma_0}
  \left\{
    \frac12\big[P(\vx,\vx,0)+P(\vy,\vy,0)\big]-P(\vx,\vy,0)
  \right\}
  \, .
\end{equation}
For example, in an infinitely long quasi-1D wire of cross-section $s$,
the solution of the diffusion equation is $P(\vx,\vy,0)=-|x-y|/(Ds)$
where $x$ is the component of $\vx$ along the wire. Hence we recover a
resistance proportional to the distance between the points,
$\clasres(\vx,\vy)=|x-y|/(s\sigma_0)$.

\section{Noise correlation function in networks of disordered wires}
\label{SectNetworks}

Let us now illustrate the calculation of the noise correlation function,
\Eq{FinalCorrFunc}, for a network of disordered wires. The main
ingredient to \Eq{FinalCorrFunc} is the solution of the diffusion
equation \EqPlain{GammaDifEq} at zero frequency. Wires allow a quasi-1D
description of diffusion, where transverse directions can be integrated
out since $P(\vx,\vy,\wo)$ is assumed to be constant on the scale of the
width of the wire.
As a result, we replace $P(\vx,\vy,\wo)\to P(x,y,\wo)/s$, where $s$ is
the cross-section of the wires and $P(x,y,\wo)$ solves the 1D diffusion
equation in the network, $x$ and $y$ being coordinates along the wires.
Recently, effective methods have been developed to solve the resulting
diffusion equation for arbitrary networks.\cite{Texier_Multiterminal_2004,
Pascaud_PersistentCurrents_1999,Akkermans_SpectralDeterminant_2000,		Texier_NetworksCylinders_2009}
We will review these methods in this Section and evaluate the 
noise correlation function for a simple example.

We start by introducing some basic notations: A network is a set of
vertices, labeled by an index $\alpha$, connected via wires of arbitrary
length, say $l_{\alpha \beta}$ for the wire connecting vertices $\alpha$
and $\beta$.
Let us define a vertex matrix $\vertexm$ as
\begin{equation}\label{DefMMatrix}
  \vertexm_{\alpha \beta}
  = \delta_{\alpha \beta}
      \sum_{\gamma} \frac{a_{\alpha \gamma}}{l_{\alpha \gamma}}
    - \frac{a_{\alpha \beta}}{l_{\alpha \beta}} \, ,
\end{equation}
where $a_{\alpha \beta}=1$ if the vertices $\alpha$ and $\beta$ are
connected and $a_{\alpha \beta}=0$ otherwise. The solution of the
diffusion equation at zero frequency between arbitrary vertices
$\alpha, \beta$ of the network is given by the entries of the inverse
matrix ${\cal M}$ divided by the diffusion constant:
\begin{equation}\label{ProbInvM}
  P(\alpha,\beta,0) = ({\cal M}^{-1})_{\alpha \beta} / D \, .
\end{equation}
This allows us to calculate
the noise correlation function between arbitrary points of a network by
inserting vertices and inverting ${\cal M}$.
As an aside, note that arbitrary boundary conditions can be
included in this scheme easily (see \Refs{Akkermans_SpectralDeterminant_2000,
Texier_NetworksCylinders_2009} for details).

\begin{figure}[t]
  \centering
	\includegraphics{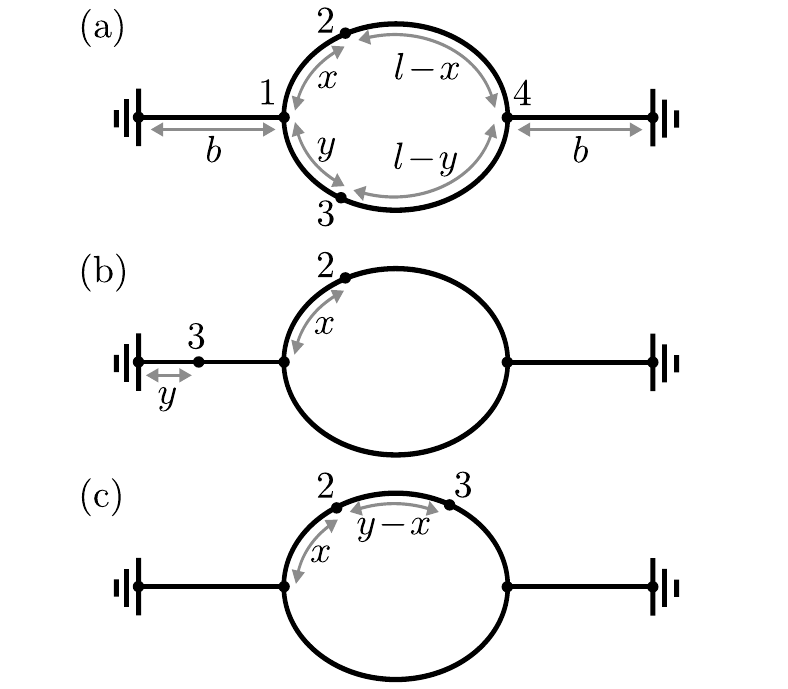}
	\caption{The network corresponding to a symmetric ring made of four
	         wires connected to two absorbing leads. The length of the
	         arcs is $l$ and the length of the connecting arms is $b$.
	         Vertices are labeled by numbers $\alpha=1,2,3,4$. Vertices
	         ``$1$'' and ``$4$'' denote the points where the ring is
	         connected to the arms. Vertex ``$2$'' is always placed in the
	         upper arc, defining the running coordinate $x$.
	         Vertex ``$3$'' determines the $y$-coordinate and is placed
	         either in the adjacent arc (panel a) or in the left arc
	         (panel b) or in the same arc (panel c).
	        }
	\label{FigGraphRing}
\end{figure}

Let us consider the network shown in \Fig{FigGraphRing}, representing a
ring connected to absorbing leads.
For simplicity, we assumed that the ring is symmetric: the two arcs are
of the same length $l$ and the connecting arms of length $b$. We
evaluate the noise correlation function for two points in this network
by inserting two vertices, called ``$2$'' and ``$3$''. Vertex ``$2$'' is
always placed in the upper arc, encoding the running coordinate $x$ in
the length of the connected wires. Vertex ``$3$'' determines the
$y$-coordinate and is placed either in the lower arc or in the left
connecting arm or in the upper arc. In the first case,
\Fig{FigGraphRing}(a), the vertex matrix,
\Eq{DefMMatrix}, is given by
\begin{equation}\label{MRing}
  {\vertexm} \!=\! \begin{pmatrix}
     \frac{1}{b}\splus\frac{1}{x}\splus\frac{1}{y} &
     -\frac{1}{x} &
     -\frac{1}{y} &
     0
   \\
     -\frac{1}{x} &
     \frac{1}{x}\splus\frac{1}{l-x} &
     0 &
     -\frac{1}{l-x}
   \\
     -\frac{1}{y} &
     0 &
     \frac{1}{y}\splus\frac{1}{l-y} &
     -\frac{1}{l-y}
   \\
     0 &
     -\frac{1}{l-x} &
     -\frac{1}{l-y} &
     \frac{1}{b}\splus\frac{1}{l-x}\splus\frac{1}{l-y}
  \end{pmatrix} .
\end{equation}
The diffusion propagator is then
given by $P(x,y,0)=(\vertexm^{-1})_{23}/D$, and we obtain from
\Eq{FinalCorrFunc} the correlation function as a function of the running
coordinates $x$, $y$ $\in[0,l]$:
\begin{align}\label{VVRingArcAdjacent}
	\CFV(x,y,\wo) = &
      \frac{\wo F(\wo)}{D\nu s}
	    \frac{b(l(2b+l) - (x+y)l + 2 x y)}{l(4b+l)} \, .
\end{align}
When vertex ``$3$'' is placed in the connecting arm, $x\in[0,l]$ and
$y\in[0,b]$ [\Fig{FigGraphRing}(b)], we get
\begin{align}\label{VVRingArcArm}
	 \CFV(x,y,\wo) =
      \frac{\wo F(\wo)}{D\nu s}
	 \frac{y(2b+l-x)}{4b+l} \, .
\end{align}
Finally, when vertex ``$3$'' is placed in the same arc of the ring as
vertex ``$2$'' [see \Fig{FigGraphRing}(c)], following the same logic we
obtain, with $0<x<y<l$,
\begin{align}\nonumber
	& \CFV(x,y,\wo) \\ \label{VVRingArcSame}
	&  = \frac{\wo F(\wo)}{D\nu s} \frac{bl(2b+l)+xl(3b+l)-ybl-xy(2b+l)}{l(4b+l)} \, .
\end{align}
All other configurations can be found by symmetry arguments.
\begin{figure}[t]
  \centering
	\includegraphics{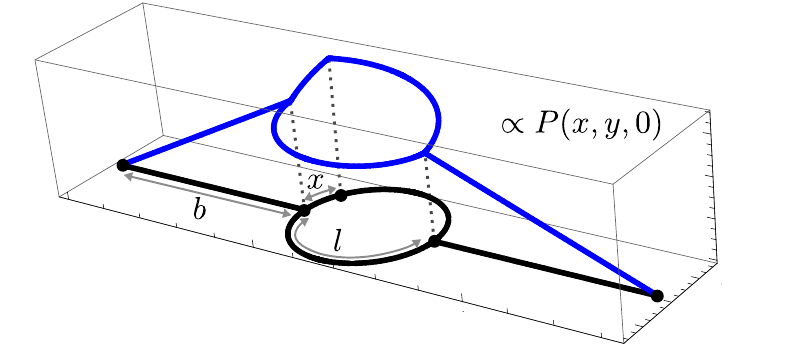}
	\caption{(color online) The solution to the diffusion equation at zero frequency,
	         $P(x,y,0)\propto\CFV(x,y,\wo)$, where $\CFV$ is given
	         by \EqsFromTo{VVRingArcAdjacent}{VVRingArcSame}, for a fixed
	         coordinate $x$ in the upper arm of the ring (indicated by the
	         dot), as a function of $y$ traversing the network.
	         $\CFV(x,y,\wo)$ is linear in $y$ and its derivative has a discontinuity
	         at $y=x$.
	        }
	\label{FluctRing}
\end{figure}
We plot $P(x,y,0)$ for $y$ traversing the whole network in
Fig.~\ref{FluctRing}. Note that the resulting function is linear in $y$
and its derivative has a discontinuity at $y=x$ (cf. \Ref{Texier_Ring_2004}).

\section{Application to dephasing}
\label{Dephasing}

The precise characterization of potential fluctuations is very important
in studying phase coherent properties of disordered metals at low
temperatures. To be specific, let us discuss a particular coherent
property: the weak localization correction to the conductivity. Let us
recall that the weak localization (WL) correction
$\Delta\sigma\equiv\overline{\sigma}-\sigma_0$ is a small contribution
to the averaged conductivity arising from quantum interference of
reversed diffusive electronic trajectories.\cite{Khmelnitskii_1984}

At low temperatures, dephasing is dominated by electron interactions,
that can be accounted for through a contribution to the phase
accumulated by two time-reversed interfering trajectories in a
fluctuating electric field~\cite{AAK_Localization_1982}:
\begin{equation}\label{PhiV} 
  \Phi[\vx(\tau)]
  =\int_0^t\dx\tau\,\big[V(\vx(\tau),\tau)-V(\vx(\tau),t-\tau)\big]
  \, .
\end{equation}
When averaged over the Gaussian fluctuations of the electric field,
$\langle e^{i\Phi}\rangle_{V}=e^{-\frac{1}{2} \langle{\Phi ^2}\rangle_{V}}$
yields a phase difference which cuts off the contributions of long
electronic trajectories. Introducing the trajectory-dependent dephasing
rate $\Gamma[\vx(\tau)]=\frac1{2t}\langle \Phi[\vx(\tau)]^2\rangle_{V}$,
the weak localization correction takes the
form:\cite{Texier_Ring_2004,Montambaux_QuasiparticleDecay_2005,
Marquardt_Decoherence_2007,Texier_NetworksCylinders_2009}
\begin{equation}\label{eq:WL}
  \Delta\sigma(\vx) = -\frac{2e^2D}{\pi}\int_0^\infty\dx t\,
  P(\vx,\vx,t)\,\langle e^{-t\Gamma[\vx(\tau)]} \rangle_{\{\vx(\tau)\}}
   \, ,
\end{equation}
where $\langle\cdots\rangle_{\{\vx(\tau)\}}$ is the average with respect
to closed diffusive trajectories of duration $t$ starting from
$\vx$ (not to be confused with the thermal average $\mean{\cdots}_{V}$
over the electric potential $V$). The phase fluctuations can then be
related to the potential fluctuations:
\begin{align}\nonumber
  &\frac12\langle \Phi[\vx]^2\rangle
  = \int_0^t\dx\tau\dx\tau'\int_{-\infty}^\infty\frac{\dx\wo}{2\pi}
  \left[ e^{-i\wo (\tau-\tau')} -  e^{-i\wo (\tau+\tau'-t)} \right] \\ \label{PhaseFluct}
  &
  \qquad \qquad \qquad \times \langle |V|^2 \rangle_\varphi(\vx(\tau),\vx(\tau'),\wo) \, .
\end{align}
Here we have introduced a new noise correlator,
\begin{equation}\label{CorrelatorPhi}
   \langle |V|^2 \rangle_\varphi(\vx,\vy,\wo)
   = \frac{1}{\nu} \, \wo F_\varphi(\wo) P(\vx,\vy,0) \, ,
\end{equation}
obtained from \Eq{FinalCorrFunc} by replacing $F(\wo)$ with a modified
function $F_\varphi(\wo)$ (given below), on the origin of which we now
comment. Equation \EqPlain{EqFDTVV} is well-known in the theory of
dephasing: its version symmetrized  with respect to frequency arises
naturally when comparing the diagrammatic calculation of the dephasing
time \cite{Aleiner_DisorderedSystems_1986,vonDelft_Decoherence_2007}
with the influence functional approach describing electrons moving in a
random Gaussian field $V$.~\cite{Chakravarty_WeakLocalization_1986,
Marquardt_Decoherence_2007,vonDelft_InfluenceFunctional_2008}
Diagrammatically, the symmetrized \Eq{EqFDTVV} represents the Keldysh
component of the screened electron interaction propagator, the only
substantial difference being that the diagrammatically calculated
correlation function involved in the dephasing process acquires
so-called ``Pauli-factors'' that account for the fact that the Fermi
sea  limits the phase space available for inelastic transitions.
\cite{vonDelft_InfluenceFunctional_2008} These factors lead to the
following replacement of the function $F(\wo)$ in \Eq{EqFDTVV} and here
also in \Eq{FinalCorrFunc}:
\begin{equation}\label{PauliBlocking}
   F(\wo)\overset{\mathrm{sym}}{\longrightarrow}
  \coth(\wo/2T) \overset{\mathrm{Pauli}}{\longrightarrow}
  \frac{\wo/2T}{\sinh^2(\wo/2T)}
  \equiv F_\varphi(\wo) \, .
\end{equation}
This restricts the energy transfer to $|\wo|<T$,
\cite{Marquardt_Decoherence_2007,vonDelft_Decoherence_2007} but does not
affect the factorization of the correlator. Inserting \Eq{CorrelatorPhi}
into \Eq{PhaseFluct} leads to
\begin{align}
  \nonumber
  &\frac12\langle \Phi[\vx]^2\rangle
  = \frac{2T}{\nu} \int_0^t\dx\tau\dx\tau'\, P(\vx(\tau),\vx(\tau'),0)
  \int_{-\infty}^\infty\frac{\dx\wo}{2\pi} 
  \\ \label{eq:3}
  &
  \qquad\qquad
  \times \left[ e^{-i\wo (\tau-\tau')} -  e^{-i\wo (\tau+\tau'-t)} \right]
  \frac{\wo}{2T}F_\varphi(\wo)
  \, .
\end{align}
The fact that the frequency dependent function
$\frac{\wo}{2T} F_\varphi(\wo)$ is symmetric allows us to add to
$P(\vx(\tau),\vx(\tau'),0)$ the term
$-\frac12\big[P(\vx(\tau),\vx(\tau),0)+P(\vx(\tau'),\vx(\tau'),0)\big]$,
which does not contribute to the integral \eqref{eq:3}. Therefore we
finally end up with the following expression for the dephasing rate,
\begin{align} \nonumber
  \Gamma[\vx(\tau)] &= e^2 T
   \int_0^t\frac{\dx\tau}{t}\int_0^t\dx\tau'\, \left[ \delta_T(\tau+\tau'-t) - \delta_T(\tau-\tau') \right] \\\label{eq:FunctionalDecoh}
  &
  \qquad \times \clasres\left(\vx(\tau),\vx(\tau')\right) \, ,
\end{align}
written in terms of the resistance $\clasres$, defined in
\Eq{Resistance}. The function $\delta_T(t)$, a broadened delta function
of width $1/T$ and height $T$, is the Fourier transform of
$\frac{\wo}{2T} F_\varphi(\wo)$, which is given by
\begin{equation}
  \delta_T(\tau)=\pi T\,w(\pi T\tau)
  \, , \quad
  w(y)=\frac{y\coth y-1}{\sinh^2y} \, .
\end{equation}
\Eq{eq:FunctionalDecoh}, which is one of the main results of our paper,
generalizes the results obtained
in \Refs{Marquardt_Decoherence_2007,Treiber_Book_2010,
Treiber_Dephasing_2009,vonDelft_InfluenceFunctional_2008} for an
infinite wire and an isolated ring to arbitrary geometry. In the
classical noise limit, $T \to \infty$, $\delta_T(\tau)$ may be replaced
by a $\delta(\tau)$ function: The second term of \Eq{eq:FunctionalDecoh}
vanishes and we recover the results of \Refs{Texier_Ring_2004,
Texier_NetworksCylinders_2009}.

Let us now illustrate \Eq{eq:FunctionalDecoh} by calculating the
dephasing time for the well-understood case of one and two-dimensional
isolated simply-connected samples. The dephasing time can be extracted
from the condition
\begin{equation}\label{eq:DefineDephasingTime}
  1\equiv \Gamma(\tau_\varphi)\tau_\varphi \, ,
\end{equation}
where $\Gamma(t)$ is
given by the functional \Eq{eq:FunctionalDecoh}, averaged over the
typical closed random walks $\vx(\tau)$ of duration $t$ in the system.
The problem is governed by the interplay of three time-scales: The
Thouless time $\tau_{\rm Th}=L^2/D$, depending on the system size $L$,
the thermal time $\tau_{T} = 1/T$ (related to the thermal length
$L_T=\sqrt{D/T}$), and the dephasing time $\tau_\varphi$.

(i) Diffusive regime, $\tau_{T} \ll \tau_\varphi \ll \tau_{\rm Th}$ ($L_T \ll L_\varphi \ll L$):
this is the regime considered in \Refs{AAK_Localization_1982,Montambaux_QuasiparticleDecay_2005}, where the
width of the broadened delta functions in \Eq{eq:FunctionalDecoh},
$\tau_{T}$, is the shortest time-scale. Thus, when averaging over paths
$\vx(\tau)$, the characteristic length-scale $|x-y|$ entering the
resistance $\clasres(x,y)$ can be determined as follows: For the first
$\delta_{T}$ term this length is governed by free diffusion, since
$|\vx(\tau) - \vx(t-\tau)| \sim \sqrt{D\tau}$, hence
$|x-y| \sim \sqrt{D\tau}$. For the second term the characteristic length
is set by the width of the delta function, $|x-y| \sim \sqrt{D\tau_{T}}$.
In 1D, where $\clasres(x,y) \sim |x-y|/\sigma_0 s$, the first term
dominates and we immediately obtain from \Eq{eq:DefineDephasingTime}
$1/\tau_\varphi \sim (e^2 \sqrt{D} T/\sigma_0 s)^{2/3}$.
In 2D, the diffuson at zero frequency is logarithmic as well as the resistance
\EqPlain{Resistance}, $\clasres(x,y) \sim \ln(|x-y|)/\sigma_0 d$, where $d$ is
the width of the sample, which can be understood from the fact that the
resistance of a plane connected at two corners scales logarithmically
with the system size.
\Eq{eq:DefineDephasingTime} gives
$1 \sim e^2 T \tau_\varphi \ln(T\tau_\varphi) / \sigma_0 d$, and for the
dephasing time:
$1/\tau_\varphi \sim e^2 T \ln(e^2/\sigma_0 d)/\sigma_0 d$.

(ii) Ergodic regime, $\tau_{T} \ll \tau_{\rm Th} \ll \tau_\varphi$ ($L_T \ll L \ll L_\varphi$):
the width of $\delta_T(\tau)$ in \Eq{eq:FunctionalDecoh} is still the
shortest time-scale but, in contrast to (i), the typical trajectories
$\vx(\tau)$ explore the whole system, setting the length-scale of
diffusion to the system size $L$, cf. \Refs{Ludwig_Ring_2004,
Texier_Ring_2004}. In full analogy to the diffusive regime, but
replacing $\sqrt{D\tau}$ by $L$, we find for 1D,
$1/\tau_\varphi \sim e^2 L T/\sigma_0 s$, and for 2D,
$1/\tau_\varphi \sim e^2 T \ln(\tau_{\rm Th}/\tau_{T})/\sigma_0 d$.\cite{Takane}
These examples show that for non-trivial geometries dephasing due to
electron interactions cannot be accounted for through a unique dephasing
rate depending only on dimensionality, but must be described by a
functional of the trajectories $\vx(\tau)$ since the qualitative
behavior of $\tau_\varphi$ follows from the geometry dependent typical
distance $|\vx(\tau) - \vx(\tau')|$.

For sufficiently low temperatures, on the other hand,
\Eq{eq:FunctionalDecoh} is capable to describe the crossover to a 0D
regime, where, apart from a dependence on the total system size,
geometry becomes unimportant: 

(iii) 0D regime, $\tau_{\rm Th} \ll \tau_{T} \ll \tau_\varphi$ ($L \ll L_T \ll L_\varphi$): here,
the width of the delta functions in \Eq{eq:FunctionalDecoh}, $\tau_{T}$,
is larger than $\tau_{\rm Th}$. Hence, the trajectories reach the
ergodic limit $\vx(\tau \geq \tau_{\rm Th}) \sim L$ before the electric
potential has significantly changed: Dephasing is strongly reduced. Let
us denote the maximal resistance reached at the ergodic limit as
$\clasres_{\rm erg}$ and replace the resistance in
\Eq{eq:FunctionalDecoh} by $\clasres \to \clasres - \clasres_{\rm erg}$,
without changing the result, since $\clasres_{\rm erg}$ is constant and
its contribution vanishes after integrating over $\tau$ and $\tau'$. The
difference $\clasres - \clasres_{\rm erg}$ is nonzero only during time
differences $\tau - \tau' \lesssim \tau_{\rm Th}$, before reaching
ergodicity. Thus, the leading contribution comes from the second
$\delta_T$ term in \Eq{eq:FunctionalDecoh}, which is constant at its
maximum $T$ during such short time-scales. We find
$1 \sim -e^2 T^2 \tau_\varphi \int_0^{\tau_{\rm Th}} \dx{\tau} [ \clasres(\vx(\tau),0) - \clasres_{\rm erg} ]$
and since the $\clasres_{\rm erg}$ term dominates, we obtain a dephasing
time $1/\tau_\varphi \sim e^2 T^2 \tau_{\rm Th} \clasres_{\rm erg}$,
independent of geometry and with the characteristic $\sim T^2$ behavior.
\cite{Sivan_QuasiParticleLifetime_1994}

\section{Conclusions}
\label{SectConclusions}
In this paper we have considered fluctuations of the scalar electric
potentials in macroscopically inhomogeneous metals. We have shown how to
relate the density fluctuations to the potential fluctuations,
emphasizing the role of electronic interactions, provided a real space
derivation of the density response function and illustrated these
general ideas for the case of networks of metallic wires. Finally we
have obtained a trajectory-dependent functional,
\Eq{eq:FunctionalDecoh}, which describes dephasing by electron
interactions for arbitrary geometries and accounts for the quantum noise
contribution. When applied to networks, \Eq{eq:FunctionalDecoh} can
describe the full crossover from the 0D to the 1D and the 2D regime.

\acknowledgments
We acknowledge illuminating discussions with F.~Marquardt, and support
from the DFG through SFB TR-12 (O.~Ye.), DE 730/8-1 (M.~T.)
and the Cluster of Excellence, Nanosystems Initiative Munich.

\begin{appendix}
\label{SectAppendix}

\section{Self consistent analysis of screening}
\label{AppendixScreening}

We recall here how to obtain \Eq{EqInitialIntEq} using a self-consistent
treatment of screening in real space.\cite{Christen_Buttiker_1996}
Starting points are the following three equations:
\noindent(i) the excess charge density is decomposed into external and induced
contributions
\begin{equation}\label{Screening1}
  \delta n(\vx,\wo)
  = n_\mathrm{ext}(\vx,\wo) + n_\mathrm{ind}(\vx,\wo) \, .
\end{equation}
(ii) The induced charge is related to the potential $V(\vx,\wo)$ by the
density response function, cf. \Eq{LinResponseV}:
\begin{equation}\label{Screening2}
  n_\mathrm{ind}(\vx,\wo)
  = -\int \dx{\vy} \, \chi(\vx,\vy,\wo) \, V(\vy,\wo) \, .
\end{equation}
(iii) The Poisson equation
\begin{equation}\label{Screening3}
  \Delta \, V(\vx,\wo) = -4 \pi e^2 \,\delta n(\vx,\wo) \, .
\end{equation}
Self-consistency lies in the fact that the response involves the
screened potential $V(\vx,\wo)$ and not the bare ``external''
potential related to $n_\mathrm{ext}(\vx,\wo)$.
The screened effective interaction between electrons
$U_{\RPA}(\vx,\vy,\wo)$ is obtained by placing an external charge at
$\vy$, so that the external density is
$n_\mathrm{ext}(\vx,\wo)=\delta(\vx-\vy)$, and associating the resulting
screened potential $V(\vx,\wo)$ in \Eq{Screening3} with
$U_{\RPA}(\vx,\vy,\omega)$. We obtain
\begin{align} \nonumber
  &  -\frac1{4\pi e^2}\Delta_\vx U_{\RPA}(\vx,\vy,\wo) 
  \\
  &
  + \int \dx{\vx'}  \, \chi(\vx ,\vx' ,\omega) \, U_{\RPA}(\vx',\vy,\omega) =\delta(\vx-\vy) \, .
\end{align}
Convolution with the Coulomb interaction gives
\Eq{EqInitialIntEq}.

\section{Current density correlations}
\label{appendix:CurrentCorrelation}

We discuss here the relation between the density and the current density
correlations. The response of the (induced) current density is
characterized by the conductivity tensor $\sigma$,
\begin{equation}
  \smean{ \hat j_\alpha(\vx,\wo) }_\mathrm{neq}
  =\int \dx{\vy} \sigma_{\alpha\beta}(\vx,\vy,\wo) \, E_\beta(\vy,\wo)
  \, ,
\end{equation}
which is related to \Eq{DefChi} by current conservation:
\begin{equation}
  \label{eq:CurrentConservation}
  \nabla_\alpha\nabla_\beta'\sigma_{\alpha\beta}(\vx,\vx',\wo)
  =-i\wo e^2\chi(\vx,\vx',\wo)
  \, .
\end{equation}
The thermal fluctuations of the current density can be obtained from
$\smean{j_\alpha j_\beta^\dagger}(\vx,\vy,\wo) = \wo F(\wo)\,\Rea{\sigma_{\alpha\beta}(\vx,\vy,\wo)}$,
in analogy to the discussion in Section~\ref{Heuristic}, assuming
time-reversal symmetry,
$\sigma_{\alpha\beta}(\vx,\vy,\wo) = \sigma_{\beta\alpha}(\vy,\vx,\wo)$.

Let us now examine the case of disordered metals. The classical
contribution to the averaged non-local dc-conductivity has been
derived in \Ref{Kane_Correlations_1988}. Their result can be generalized
straightforwardly to non-zero frequencies,
\begin{equation}\label{eq:NonLocalSigma}
  \overline{\sigma}_{\alpha\beta}(\vx,\vx',\wo)=
  \sigma_0\,
  \left[
    \delta_{\alpha\beta}\delta(\vx-\vx')
    - D \nabla_\alpha\nabla_\beta'
    P(\vx,\vx',\wo)
  \right]
  \, ,
\end{equation}
which obeys the
condition \EqPlain{eq:CurrentConservation} with \Eq{ChiBar} substituted
for $\chi$. For the current correlations, we find
\cite{Footnote_UCFvsThermal}
\begin{align}
   \overline{ \smean{j_\alpha j_\beta^\dagger}(\vx,\vx',\wo) }
  =& \sigma_0\,\wo F(\wo)\,
  \big\{
    \delta_{\alpha\beta}\delta(\vx-\vx')
   \nonumber \\
    &- D \nabla_\alpha\nabla_\beta'
  \Rea{P(\vx,\vx',\wo)}
  \big\}
  \, .
\end{align}
Since the diffuson, $P(\vx,\vx',\wo)$, decays exponentially on a length
scale $L_\wo=\sqrt{D/\wo}$, this expression shows that current
correlations can be considered as purely local over the scale
$||\vx-\vx'||\gg L_\wo$, i.e
$\overline{ \smean{|j|^2}(\vx,\vx',\wo) }\simeq\sigma_0\wo F(\wo)\delta(\vx-\vx')$.
In the limit of classical noise, $F(\wo)\,\wo\simeq2T$, we recover
precisely \Eq{Noisej}.

\end{appendix}


\begin{thebibliography}{30}
\bibitem{Footnote_VOperator}
In general, if the electromagnetic field is not quantized, as is the
case here, the potential is not an operator; the assumption of a
potential operator arises from electronic interactions that relate the
internal electronic density operator to the electric potential.
\end{thebibliography}
\end{document}